\documentclass[10pt,
               aps,
               pre,
               twocolumn,
               showpacs,
               ]{revtex4-2}
\usepackage[utf8]{inputenc}
\usepackage{graphicx}
\usepackage{xcolor}
\graphicspath{{images_dia_to_para/}}
\usepackage[caption=false]{subfig}
\usepackage{amsmath, amsfonts, amssymb}
\usepackage{hyperref}

\makeatletter
\newcommand*{\rom}[1]{\expandafter\romannumeral #1}
\makeatother

\begin{document}
\title{Orbital Magnetism of Active Viscoelastic Suspension}
\author{M Muhsin}
\affiliation{Department of Physics, University of Kerala, Kariavattom, Thiruvananthapuram-$695581$, India}

\author{M Sahoo}
\email{Corresponding author :  jolly.iopb@gmail.com}
\affiliation{Department of Physics, University of Kerala, Kariavattom, Thiruvananthapuram-$695581$, India}

\author{Arnab Saha}
\email{Corrresponding author : sahaarn@gmail.com\\}
\affiliation{Department of Physics, University of Calcutta, 92 Acharya Prafulla Chandra Road, Kolkata-700009, India}
\date{\today}

\begin{abstract}
\begin{description}
\item[Abstract] {\textcolor{black}{We consider a dilute suspension of active (self-propelling) particles in a visco-elastic fluid. Particles are charged and constrained to move in a two dimensional harmonic trap. Their dynamics is coupled to a constant magnetic field applied perpendicular to their motion via Lorentz force. Due to the finite activity, the generalised fluctuation-dissipation relation (GFDR) breaks down, driving the system away from equilibrium. While breaking GFDR, we have shown that the system can have finite classical orbital magnetism only when the dynamics of the system contains finite inertia. The orbital magnetic moment has been calculated exactly. Remarkably, we find that when the elastic dissipation time scale of the medium is larger (smaller) than the persistence time scale of the self-propelling particles, the system is diamagnetic (paramagnetic). Therefore, for a given strength of the magnetic field, the system undergoes a novel transition from diamagnetic to paramagnetic state (and vice-versa) simply by tuning the time scales of underlying physical processes, such as, active fluctuations and visco-elastic dissipation. Interestingly, we also find that the magnetic moment, which vanishes at equilibrium, behaves non-monotonically with respect to increasing persistence of self-propulsion, that drives the system out of equilibrium}}
\end{description}
\end{abstract}

\maketitle

\section{INTRODUCTION}

{\textcolor{black}{Inertia can have profound effect on dynamics. It can be a system of particles forming a rigid body or a fluid, inertia can be equally important in the dynamics of both. From celestial bodies to a spinning top of everyday life, it is evident that the effect of inertia is ubiquitous in rigid body mechanics. In case of fluids, at high Reynolds number \cite{frisch1995turbulence}, inertia plays important role from simpler problems of fluid mechanics, such as, inviscid flows, potential flows, laminar flows etc. all the way to one of the most challenging problems - turbulence 
\cite{landau1959fluid, frisch1995turbulence}.\\}}
Though, with respect to the inertial dynamics mentioned above,  the dynamics in the world of motile micro-organisms (e.g. bacteria, green algae, sperm cells, white blood cells and red blood cells or even smaller scale objects like motor proteins etc.) and synthetic micro-swimmers (e.g. active colloids \cite{howse2007self}), are fundamentally different \cite{lauga2009hydrodynamics, bechinger2016active}. First of all, unlike the passive particles, the micro-organisms can self-propel, consuming energy from their surrounding. They spontaneously generate flow into the system, driving it far from equilibrium \cite{dombrowski2004self}. Secondly, because of their size limitation and the highly viscous medium in which the micro-organisms self-propel, the typical Reynolds number is around $10^{-4}$ or even lesser \cite{lushi2014fluid, purcell1977life}. The typical time required for the micro-swimmers at low Reynolds number to dissipate their momenta is around $10^{-7}$ seconds \cite{purcell1977life}. Therefore, for all practical purpose, through out their journey, the momentum of such swimmer remains constant over time, Consequently, the inertia has negligibly small effect to the dynamics of such self-propellers. Most of the research on $\it {active}$ systems so far is done in this low Reynolds number limit where one consider Stokes flow including active stresses (namely, active hydrodynamics \cite{marchetti2013hydrodynamics}) and/or over-damped  Brownian motion including self-propelling forces and torques \cite{bechinger2016active}.\\

Clearly, if we push the envelop further considering larger self-propelling objects moving in a medium with lower viscosity, inertial effect will become prominent. Typically, active particles moving in a low viscosity media starting from millimeter-sized (and onwards), are strongly influenced by inertial forces and torques. Macroscopic particles of granular material with in-built self-propelling or self-vibrating mechanisms  (e.g. internal vibration motor, vibrating plate, mini-robots etc. \cite{narayan2007long, weber2013long, scholz2018rotating, deblais2018boundaries, kudrolli2008swarming,deseigne2010collective, patterson2017clogging, junot2017active, notomista2019study, mayya2019non, klotsa2019above}), macroscopic swimmers \cite{gazzola2014scaling, saadat2017rules, gazzola2015gait, shahsavan2020bioinspired}, flying insects \cite{sane2003aerodynamics} are apt examples where inertia can play a significant role in their dynamics, both in the single particle level as well as in the collective level. Recently there are studies from theoretical as well as experimental perspectives, focusing on inertial effects of self propelled particles \cite{dauchot2019dynamics, walsh2017noise, lanoiselee2018statistical, scholz2018inertial, mijalkov2013sorting, gupta2020flocking}. Self propelling robots can be fabricated in the macroscopic length scales \cite{leyman2018tuning, mijalkov2016engineering} to explore the inertial effects. Active Langevin model including inertia can describe the dynamics of inertial self-propellers well \cite{caprini2021inertial, sprenger2021time, caprini2021spatial}. It is observed that by fine tuning inertia, some of the fundamental properties of active systems are qualitatively modified. For example, it has been shown experimentally as well as theoretically that inertia can induce delay between the orientation dynamics and velocity of active particles which has profound influence in their long-time dynamics \cite{scholz2018inertial}. In the presence of inertia, different dynamical states are developed by self-propelling particles confined in a trap. The transition between these dynamical states is continuous or discontinuous, crucially depends on the inertia present in the system \cite{dauchot2019dynamics}. One of the fundamental feature of active Brownian systems is motility induced phase separation (MIPS). It is strongly influenced and suppressed by the presence of inertia. It has also been shown that due to inertia, the coexisting phases of high and low particle density, obtained by MIPS, have widely different kinetic temperatures, which is in contrast with equilibrium phase separation \cite{mandal2019motility}.  
 
Being motivated by the above recent findings on inertial active systems, here we will report the magnetic properties of a dilute suspension of charged active particles under external magnetic field.  In particular we will show that if the particles posses finite inertia, only then such system can exhibit orbital magnetism and goes through a transition from the paramagnetic to the the diamagnetic state in different regime of parameter space of the model. The transition between paramagnetic and diamagnetic states depends crucially on the interplay of different time scales related to the physical processes (e.g. active fluctuations and dissipation) involved in the dynamics of the system. 

Before going into the details of our findings, we will introduce here the medium in which the active particles are suspended. We consider the medium to be viscoelastic with transient elasticity. The elastic forces exerted by the medium to the self-propelling particles dissipate within a finite time beyond which, particles are dragged only by the viscous forces. Theoretically one can also consider the viscous limit of the problem where elastic forces dissipate very fast and only viscous forces are left to drag the suspended particles. In experiments, usually polymers are added to the viscous fluids to make it transiently viscoelastic\cite{saad2019diffusiophoresis}. Viscoelasticity can Trigger fast transitions of a Brownian particle in a double well optical potential\cite{ferrer2021fluid}. It can add remarkable features to the dynamics of the active system suspended in it. For example, it enhances rotational diffusion of the active particles\cite{gomez2016dynamics, qi2020enhanced}. In a viscoelastic environment, the self-propelling colloids exhibit transition from enhanced angular diffusion to persistent rotational motion beyond a critical propulsion speed \cite{narinder2018memory}. Viscoelasticity can enhance or retard the swimming speed of a helical swimmer, depending on the geometrical details of the swimmer and the fluid properties \cite{spagnolie2013locomotion}. Natural active systems are often found in viscoelastic environment. It is imperative to study viscoelastic effects on their dynamics \cite{shen2011undulatory, saha2016determining}. Theoretically it has been shown that, elasticity in a non-Newtonian fluid  can suppress cell division and cell motility \cite{emmanuel2020active}. In viscoelastic environment active pulses can reverse the flow \cite{plan2021activity}. In case of chemically powered self-propelling dimers, fluid elasticity enhances translational and rotational motion at the single particle level whereas in the multi-particle level, it enhances alignment and clustering\cite{sahoo2021role}.            

In this article we consider a dilute viscoelastic suspension of inertial, self-propelling (thereby, active) particles in two dimensions (2D). The suspension is confined in a 2D harmonic potential. The particles are charged and subject to a constant magnetic field perpendicular to the plane in which the particles are moving \cite{jayannavar2007charged, saha2008nonequilibrium, jayannavar1981orbital,kumar2012classical}. Since the system is dilute, the inter-particle interactions are considered to be negligible in comparison to other forces present in the system. We will focus on how such system respond to the presence of external magnetic field. We quantify the response by evaluating the magnetic moment of the system and its characteristic features with respect to the time scales signifying various physical processes occurring within the system, such as : (1) time scale related to the correlation of active fluctuations (thanks to self propulsion) and (2) time scale related to dissipation. The time scale associated with active fluctuations originate from the persistence of the self-propellers to move along a certain direction despite random collision from surrounding fluid particles. The time scale associated with dissipation signifies the characteristic time of the surrounding viscoelastic fluid {\it within} which the elastic dissipation dominates and {\it beyond} which the viscous dissipation dominates. Note that here we consider the elastic dissipation being transient, decays within a finite time whereas the viscous dissipation prevails for large time. When these two time scales are equal, the generalised fluctuation-dissipation relation (GFDR) \cite{kubo1966fluctuation} holds and the system remains non-magnetic with zero magnetic moment. Conversely, when the fluctuation and dissipation time scale are unequal, GFDR breaks down and the system shows non-zero magnetic moment under the influence of external magnetic field. Depending on these two competing time scales, our analysis reveals that the suspension can exhibit either paramagnetic or diamagnetic behaviour. In particular,  we show that when the persistence in active fluctuation dominates over the dissipation, the system manifests paramagnetism and when the dissipation dominates over the self-propulsion, the system exhibits diamagnetism. Therefore,  fixing the external magnetic field to a non-zero constant value, when the time scales of these two physical processes, namely, active fluctuations and dissipations (thanks to elasticity and viscosity of the medium) are tuned, the system undergoes a transition from diamagnetic states to paramagnetic states and vice-versa. Importantly, it has also been shown that all these magnetic characteristics of the system crucially depend on the presence of inertia. If inertia of the system is negligibly small in compared to the dissipative as well as active forces, all the existing forces in the system cancel each other, providing a zero net-force, and then the magnetic moment vanishes. As a result the system looses its magnetic characteristics.  

In the next section, we address the model by which the problem is described in detail. The results are systematically detailed in the subsequent sections and finally we conclude.  

\section{MODEL AND METHOD}
\label{sec:model}

\underline{Model}: We consider $N$ non-interacting active (self-propelling) particles suspended in the viscoelastic medium at temperature $T$. The particles are at positions ${{\bf r}}_i(t) = x_i(t)\hat x+y_i(t)\hat y$  and at velocities $\dot{\bf r}_i={\bf v}_i$ at time $t$, where $i=1,2,3,....N$ is the particle index, $(\hat x, \hat y)$ are the unit vectors along $X$ and $Y$. Each particle has charge $|q|$ and constrained to move on $X$-$Y$ plane. Particles are confined within a 2D harmonic trap, $U(x_i,y_i) = \frac{1}{2}k(x_i^2+y_i^2)$ where $k$ is the spring-constant. The particles are also subjected to an external constant magnetic field ${\bf B}=B\hat z$ where $\hat z$ is the unit vector along $Z$. The equation of motion of $i$-th particle is  given by, 

\begin{equation}
m\ddot {\bf r}_i=-\gamma\int^t_0 g(t-t^{\prime}){\bf v}_i(t^{\prime})dt^{\prime} + \frac{|q|}{c}({\bf v}_i\times {\bf B})-k{\bf{r}}_i+{\sqrt{D}{\boldsymbol \xi}_i(t)}.
\label{model}
\end{equation}
Here $m$ is the mass of the particle and $\dot{\bf r}_i={\bf v}_i$ is  its velocity.  To take inertia into account, in the equation of motion of the particle ( Eq.{\eqref{model}}), we consider $\ddot{\bf r}_i=\dot{\bf v}_i$ as the acceleration of the particle.  

As the particles are self-propelling, despite the random collision of the particles of the surrounding viscoelastic fluid, they remain persistent to move along a certain direction up to a finite time scale. Moreover the dynamics of the particles contains finite memory due to the elasticity of the fluid and therefore the dynamics is non-Markovian \cite{goychuk2012viscoelastic}. The first term in RHS of Eq.~{\eqref{model}} represents the drag force on the particle because of the friction with the surrounding medium. Due to the elasticity present in the medium, the drag at time $t$ not only depends on the velocity of the particle at that particular time $t$, rather it depends on the weighted sum of all the past velocities within the time-interval between $0$ and $t$. As we consider the time-evolution of the particles to be stationary, the weight function $g$ should be a function of $(t-t')$, with $t \geq t'$. In particular we choose the weight-function (in other words, friction kernel) as 
\begin{equation}
g(t-t') = \frac{1}{2t_c^\prime}e^{-\frac{(t-t^\prime)}{t_c^\prime}}  \phantom {xxxxx}  t \geq t'
\label{eq:exponential_gamma}
\end{equation}
The above kernel gives the maximum weight to the current velocity ${\bf v}_i(t)$, whereas the weight to the past velocities decays exponentially with the rate $1/t_c'$. The time $t\sim t_c'$ is the time required for elastic dissipation to decay substantially. Therefore, for time $t > t_c'$, the viscous dissipation dominates. In the limit $t_c'\rightarrow 0$, $g(t-t')=\delta(t-t')$ and consequently the system is left with only viscous dissipation. The friction kernel $g(t-t')$ captures the Maxwellian viscoelasticity formalism, where at large enough time the fluid becomes viscous through a transient viscoelasticity, allowing the elastic force to relax down to zero \cite{goychuk2012viscoelastic}.

The second term in the RHS of  Eq.~{\eqref{model}} represents Lorentz force \cite{maxwell1873treatise} caused by the magnetic field which couples $X$ and $Y$.  The third term  in the RHS of  Eq.~{\eqref{model}} represents the harmonic confinement. The term with ${\boldsymbol{\xi}}$ appeared in Eq.~\eqref{model} represents active, coloured (and thereby athermal) noise. The moments of $\xi_{\alpha}(t)$ are given by,

\begin{equation}
\langle \xi_\alpha(t) \rangle = 0, \phantom{xx}\\
\langle \xi_\alpha(t)\xi_\beta(t^\prime) \rangle = 
\frac{\delta_{\alpha\beta}}{2t_{c}}e^{-\frac{|t-t^\prime|}{t_c}}
\label{eq:noise_coor}
\end{equation}
where $(\alpha,\beta)\in (X,Y)$. Here $t_c$ is the noise correlation time. The effective noise in the dynamics has finite correlation and it decays exponentially with time constant $t_c$, representing the persistence of the self-propellers. Up to $t=t_c$, the self-propellers remain quite persistent to move along a direction same as its previous steps, despite making random collisions with surrounding fluid particles. When $t > t_c$,  change of the direction with respect to the previous step becomes more probable. In the limit of $t_c\rightarrow 0$, with $D=2\gamma K_BT$, the active fluctuations become thermal and the system becomes passive. Therefore, in the current model for active (self-propelling) particles, finite and non-zero $t_c$  quantifies the activity of the system. Thus we model the dynamics of the particle as an Active Ornstein-Uhlenbeck Process (AOUP)\cite{fodor2016far, martin2021statistical}. But with inertia and elastic forces from the medium, the dynamics can be represented by the generalized Langevin's equation \cite{goychuk2012viscoelastic}. One may note that when $t_c=t_c^{\prime}$, we get $\langle \xi_{\alpha}(t)\xi_{\beta}(t')\rangle =\delta_{\alpha\beta}g(t-t^{'})$, which is the generalized fluctuation dissipation relation (GFDR).  Here we will explore both the situations where GFDR holds and where it does not hold ( i.e. both for $t_c=t_c'$ and $t_c \neq t_c'$). A special case where the elasticity of the fluid dissipates very fast compared to the active fluctuations (i.e. for $t_c' \rightarrow 0$ and $t_c > 0$), will be discussed in greater detail.

\underline{Method}: Now we solve the model Eq.~\eqref{model} to evaluate the magnetic moment of $i$-th particle, ${\bf M}_i=\frac{|q|}{2c}{\bf r}_i\times {\bf v}_i$ in steady states. Introducing the complex variable $z_i=x_i+jy_i$ $(j=\sqrt{-1})$, we rewrite Eq.~\eqref{model} as,

\begin{equation}
\ddot{z}_i(t) + \int\limits_0^t \Gamma g(t-t')\dot{z}_i(t^\prime) \, dt^\prime - j\ \omega_c\dot{z}_i(t) + \omega_0^2z_i(t) = \varepsilon_i(t)
\label{eq:model_complex}
\end{equation}
where the parameters are: $ \Gamma=\frac{\gamma}{m}, \omega_c = \frac{|q| B}{m c},\quad \omega_0 = \sqrt{\frac{k}{m}} $ and $\varepsilon_i(t) = \sqrt\frac{2\Gamma K_BT}{m}\left(\xi^x_i(t) + j\ \xi^y_i(t)\right)$.

 By performing the Laplace transform of the complex variable $z_i(t)$ and $\dot{z_i}(t)$ we get:  $\mathcal{L}\{z_i\}(s) = \int\limits_0^\infty e^{-s t} z_i(t)\, dt $ and $\mathcal{L}\{\dot{z}_i\}(s) = s\,\mathcal{L}\{z_i\}(s)$, where

\begin{equation}
\mathcal{L}\{z_i\}(s) = \frac{\left(\frac{1}{t_c^\prime} + s\right)\mathcal{L}\{\varepsilon_i\}(s)}{s^3 + \left(\frac{1}{t_c^\prime} - i\omega_c\right) s^2 + \left(\frac{\Gamma}{t_c^\prime} - \frac{i\ \omega_c}{t_c^\prime} + \omega_0^2 \right)s + \frac{\omega_0^2}{t_c^\prime}} 
\label{eq:model_laplace}
\end{equation}

The denominator in the Eq.~\eqref{eq:model_laplace} can be factorized as follows:
\begin{align}
D &= (s - s_1)(s-s_2)(s-s_3)
\label{eq:denominator}
\end{align}
The roots of Eq~\eqref{eq:denominator} can be found using Cardano's method. Substituting Eq.~\eqref{eq:denominator} in Eq.~\eqref{eq:model_laplace} and using method of partial fraction,

\begin{widetext}
\begin{subequations}
\begin{align}
\mathcal{L}\{z_i\}(s) &= \sum_{k=1}^3\frac{a_k}{s-s_k}\mathcal{L}\{\varepsilon_i\}(s)
\label{eq:series_solution_a}\\[0.5em]
\mathcal{L}\{\dot{z_i}\}(s) &= \sum_{l=1}^3\frac{b_l}{s-s_l}\mathcal{L}\{\varepsilon_i\}(s)
\label{eq:series_solution_b}
\end{align}
\end{subequations}
\noindent
where $(a_i,b_i)$'s are found solving,

\begin{eqnarray}
\nonumber
&& a_1+a_2+a_3=0; \phantom{xx} a_1(s_2+s_3)+a_2(s_3+s_1)+a_3(s_2+s_1)=-1,\phantom{xx}
a_1s_2s_3+a_2s_3s_1+a_3s_2s_1=\frac{1}{t_c^{\prime}}\\
&& b_1+b_2+b_3=1; \phantom{xx}\nonumber b_1(s_2+s_3)+b_2(s_3+s_1)+b_3(s_2+s_1)=-\frac{1}{t_c^{\prime}},\phantom{xx} b_1s_2s_3+b_2s_3s_1+b_3s_2s_1=0
\end{eqnarray}
\end{widetext}

By taking inverse Laplace transform of Eq.~\eqref{eq:series_solution_a} and Eq~\eqref{eq:series_solution_b}, we get $z_i(t)$ and $\dot{z}_i(t)$ as 

\begin{subequations}
\begin{align}
z_i(t) = \sum_{k=1}^3 a_k \int\limits_0^t e^{s_k(t-t^\prime)}\varepsilon_i(t^\prime)\,dt^\prime \\[0.5em]
\dot{z_i}(t) = \sum_{l=1}^3 b_l \int\limits_0^t e^{s_l(t-t^{\prime\prime})}\varepsilon_i(t^{\prime\prime})\, dt^{\prime\prime}
\end{align}
\label{eq:solution}
\end{subequations}
The average orbital magnetic moment of the particle is rewritten as
\begin{equation}
\langle {\bf M(t)} \rangle = \frac{|q|}{2c}\langle |{\bf r}_i\times{\bf v}_i|\rangle\hat z=M\hat z=\frac{|q|}{2c} \mathrm{Im}{\left(\langle z_i(t)\dot{z}_i^*(t) \rangle\right)}\hat z
\label{mag_mom}
\end{equation}
Here, $\mathrm{Im}{(\dots)}$ denotes the imaginary part, and `$*$' denotes the complex conjugation.  Using Eq.~\eqref{eq:solution}, the average orbital magnetic moment is calculated to be

\begin{widetext}
\begin{equation}
M(t) = \frac{|q|\Gamma K_B T}{m\ c\ t_c} \mathrm{Im}{\left( \sum_{k, l=1}^3 a_k\,b_l^* \int\limits_0^t\, \int\limits_0^t e^{s_k(t-t^\prime)}e^{s_l^*(t-t^{\prime\prime})} e^{-\frac{|t^\prime - t^{\prime\prime}|}{ t_c}}\, dt^\prime\, dt^{\prime\prime} \right)}
\end{equation}
Since we are interested in the time asymptotic limit or steady state, by straightforward integration in the long time ($t\rightarrow\infty$) limit, we obtain the average orbital magnetic moment as:

\begin{equation}
 M(\infty)  = \frac{|q|\Gamma K_B T}{m\ c\ t_c} \mathrm{Im}{\left( \sum_{k, l=1}^3 a_k\,b_l^* \left[ \frac{ t_c^2}{(1+s_i t_c)(-1+s_j^* t_c)} + \frac{2 t_c}{(s_i + s_j^*)(-1+s_i^2 t_c^2)} \right] \right)}
\label{eq:mag_mom_integral}
\end{equation}
\end{widetext}
\begin{figure*}[!ht]
    \includegraphics[scale=0.45]{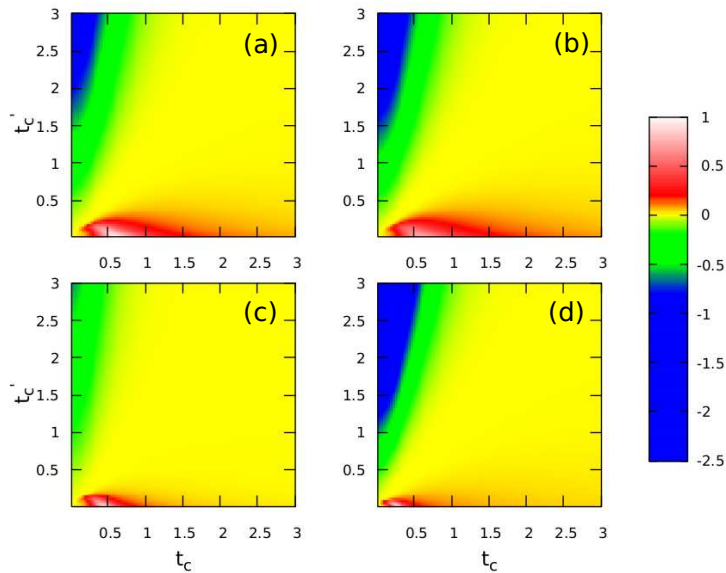}
\caption{$\langle M(\infty)\rangle$ [Eq.~\eqref{eq:mag_mom_integral}] as a function of $t_c$ and $t_c^\prime$; (a) with $\omega_c=1$, $\omega_0 = 1$, and $\Gamma = 1$, (b) with $\omega_c = 1.5$, $\omega_0 = 1$, $\Gamma = 1$ (c) with $\omega_0 = 1.5 $, $\omega_c = 1$, $\Gamma = 1$, (d) with $\Gamma = 3$, $\omega_0 = 1$, $\omega_c = 1$. Both $K_BT$ and $\frac{|q|}{mc}$ are assumed to be unity here. }
\label{Moment}
\end{figure*}

\section{Discussion}
In Fig.{\ref{Moment}} we have plotted $M$ as a function of $t_c$ and $t_c^{\prime}$ with different values of $(\omega_0, \omega_c, \Gamma)$. From this exact result, the following magnetic features of the system become apparent:

(a) First of all, along the diagonal $t_c'=t_c$, where GFDR holds, the average orbital magnetic moment $M=0$. Therefore, along the diagonal, the system becomes non-magnetic even if the dynamics contains memory. It is the reminiscent of the exact result stating that the thermal average of magnetisation of any classical system in equilibrium is zero \cite{van1932theory}.

(b)On $t_c'- t_c$ plane where $t_c' > t_c$, the system becomes diamagnetic as $M < 0$. In this regime (elastic) dissipative forces dominate over the self-propulsion. Moreover, the externally applied magnetic field induces a magnetic field within the system of charged particles which opposes the external magnetic field itself. Therefore the system is diamagnetic when $t_c' > t_c$.

However on $t_c-t_c'$ plane where $t_c > t_c'$, the self-propulsion persists for longer time as compared to the elastic dissipation of the particles. As a result the induced magnetic field within the system follows the external magnetic field. Hence the system becomes paramagnetic. 

Therefore, for a given strength of the external magnetic field, the system exhibits a transition from the diamagnetic state to the paramagnetic state and vice-versa simply by tuning the rate of the dissipation and the persistence time scale of self-propulsion. Moreover, if the direction of the magnetic field is reversed, the system shows opposite behaviour regarding the phases.

(c) While passing from diamagnetic states to the paramagnetic states or vice-versa,  the system undergoes a considerably large regime of non-magnetic states with $M=0$ on the $t_c$-$t_c^{\prime}$ plane. This regime includes the diagonal $t_c'=t_c$. However, it also includes a regime across the diagonal where $t_c\neq t_c'$ (but they are comparable). It is to be noted that across this regime, the magnetization of the system still remains zero. This regime grows as one proceeds along the diagonal. It confirms that if GFDR holds good, it implies that the average orbital magnetic moment is always zero, however the converse is not true. Average magnetic moment can still be zero even if the system is driven out of equilibrium where GFDR does not hold good.  

Similarly, we also note that when $t_c>>t_c'$, GFDR is obviously broken and the system is far from equilibrium conditions. Remarkably, in this regime, the paramagnetic moment get reduced further and approaches to zero. Hence the system becomes non-magnetic. This is because for large enough $t_c$, due to high persistence in the dynamics, the self propulsion of the particles overcomes the influence of the external magnetic field and induces a persistent rectilinear motion into them. This hinders the particles to form closed current loops which are essential to exhibit finite magnetic moment \cite{jackson1999classical}.  

(d) Apart from the fluctuation and dissipation time scales, the parameters $\Gamma, \omega_0$ and $\omega_c$ can also alter the profile of $M(t_c,t_c^{\prime})$. With increasing $\omega_c$, $M$ increases up to a maximum value beyond which $M$ decreases with further increase in $\omega_c$. Therefore $M$ has a non-monotonic dependence on $\omega_c$. The other two parameters, namely $\Gamma$ and $\omega_0$ can only reduce $M$ monotonically.

For further insight, we consider the following special case where the friction kernel is a delta function (note that in the limit $t_{c}^\prime \rightarrow 0$,  the exponential friction kernel Eq~\eqref{eq:exponential_gamma} becomes delta function) and the noise correlation is still exponential with correlation time, $t_{c}$. In this limit, only viscous dissipation takes place and the equation of motion [Eq~\eqref{model}] reduces to

\begin{equation}
    m \ddot{z}_i = -\gamma \dot{z}_i + j\frac{|q| B}{c}\dot{z}_i - k z_i + \varepsilon_i(t),
    \label{eq:model_cart_limit}
\end{equation}
where $j=\sqrt{-1}$. This dynamics can be considered as inertial active Ornstein Uhlenbeck process (IAOUP). The over-damped version of which, namely active Ornstein Uhlenbeck process (AOUP) is now commonly used to represent over-damped motion of active Brownian particles and successfully used to describe important features like MIPS \cite{fodor2016far} and dynamical heterogeneities of active systems at high densities \cite{nandi2018random}. We have exactly solved the dynamics both analytically as well as using computer simulation. Following the similar procedure as before, the solution of the dynamics Eq~\eqref{eq:model_cart_limit}, $z_i(t)$ and $\dot{z}_i(t)$ can be obtained as
\begin{subequations}
\begin{align}
z_i(t) = \sum_{k=1, 2} a_k \int\limits_0^t e^{s_k(t-t^\prime)}\varepsilon_i(t^\prime)\,dt^\prime \\[0.5em]
\dot{z}_i(t) = \sum_{l=1, 2} b_l \int\limits_0^t e^{s_l(t-t^{\prime\prime})}\varepsilon_i(t^{\prime\prime})\, dt^{\prime\prime}.
\end{align}
\label{eq:solution_limit}
\end{subequations}
Where, $s_k$'s are given by 
\begin{subequations}
\begin{align}
s_{(1,2)} &= \frac{1}{2}\left(-\left(\Gamma - i\ \omega_c\right) \pm \sqrt{\left(\Gamma - i\ \omega_c\right)^2 - 4\omega_0^2}\right) \\[0.5em]
\nonumber
\end{align}
\label{eq:si_limit}
\end{subequations}
and $a_k$'s and $b_l$'s are given by

\begin{subequations}
\begin{align}
a_1 &= \frac{1}{s_1 - s_2},\quad a_2 = -a_1 \label{eq:ai_limit} \\[0.5em]
b_1 &= \frac{s_1}{s_1 - s_2},\quad b_2 = -\frac{s_2}{s_1}b_1 \label{eq:bi_limit}
\end{align}
\end{subequations}
Using the solutions in Eq~\eqref{eq:solution_limit} and Eq[\ref{mag_mom}], the average magnetic moment in the long time limit is given by 
\begin{align}
M_r & = M(\infty)_{t_c^\prime \rightarrow 0} \nonumber \\
    & = \frac{|q| K_B T}{m c} \left[ \frac{ t_c^2\omega_c}{\left(\left(1+ t_c\left(\Gamma+ t_c\omega_0^2\right)\right)^2 +  t_c^2\omega_c^2\right)} \right]
\label{MagMomR}
\end{align}

It is evident from the aforementioned exact result in Eq.{\ref{MagMomR}} that for $t_c=0$, $M_r=0$ which is consistent with equilibrium result, namely Bohr-van Leeuwen theorem (BvL \cite{van1932theory}). When $t_c > 0$, the system goes away from equilibrium and therefore, it is intuitive that $M_r$ also shoot up. When $t_c > 0$ but small, $M_r=\left(\frac{|q|K_BT\omega_c}{mc}\right)t_c^2$, implying that in this limit $M_r$ increases with $t_c^2$.\\

On the other hand, It is also evident from the aforementioned exact result in Eq.{\ref{MagMomR}} that for $t_c\rightarrow \infty$, $M_r=0$. From the point of view of equilibrium physics, it is counter intuitive. This is because in this limit the system is very far from equilibrium and hence the equilibrium result ($M_r=0$) should not hold good. However, on a closer inspection one may note that in this limit, the persistence of the self-propulsion is so high that it is not allowing the particles to form a close current loop under the influence of the magnetic field, which is essential to build up finite magnetic moment within the realm of classical equilibrium physics. Therefore the magnetic moment vanishes in $t_c \rightarrow \infty$ limit. But for finite and large $t_c$, $M_r=\left(\frac{|q|K_BT\omega_c}{mc\omega_0^4}\right)t_c^{-2}$ signifying that for large $t_c$, $M_r$ decreases with $t_c^{-2}$.\\
Taking together, we find $M_r$ has a non-monotonic dependence on $t_c$. We exactly determine the optimum $t_c$ $(\equiv \tau_0)$ at which $M_r$ is maximum by solving 

\begin{equation}
\frac{d M_r}{d t_c} \bigg |_{ t_c=\tau_0} = \omega_0^4 \tau_0^4 + \Gamma \omega_0^2 \tau_0^3 - \Gamma \tau_0 - 1 = 0.
\end{equation}

Here $\tau_0=\frac{1}{\omega_0}$ solves the equation, which implies the optimum $t_c$ or $\tau_0$ is independent of $\omega_c$ (i.e. independent of  magnetic field), but it is inversely proportional to $\omega_0$ (i.e. varies inversely with the strength of the harmonic trap). The value of the magnetic moment at $\tau_0$ is given by $M_r(\tau_0)=\frac{|q|K_BT}{mc}\left[\frac{\omega_c}{\omega_c^2+(\Gamma+2\omega_0)^2}\right]$. Clearly $M_r(\tau_0)$ can only decrease with $\omega_0$ whereas it is non-monotonic with $\omega_c$ (For small $\omega_c$, $M_r(\tau_0)$ increases with $\omega_c$ but for large enough $\omega_c$ it decreases and finally $\lim_{\omega_c\rightarrow \infty} M_r(\tau_0)=0$).  

\begin{figure}[!hb]
    \includegraphics[scale=0.3]{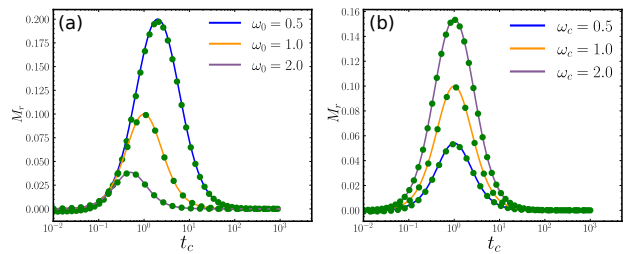}
\caption{ $M_r$ [Eq.~\eqref{MagMomR}]  is plotted with $t_c$ (at $t_c'\rightarrow 0$ limit) from analytics as well as from numerical simulation (green dotted lines), for (a) various values of $\omega_0$, keeping $\omega_c$ and $\Gamma$ to be fixed as unity (b) for various values of $\omega_c$ , keeping $\omega_0$ and $\Gamma$ as fixed to unity. Both $K_BT$ and $\frac{|q|}{mc}$ are also assumed to be unity here.  }
\label{MR}
\end{figure}

In figure [\ref{MR}] we show the behaviour of $M_r$ with $t_c$ for different values of $\omega_0$ and $\omega_c$ both from theoretical calculation as well as from simulation. Numerical simulation of the dynamics has been carried out using Heun's method algorithm. We perform the simulation with a time step of 0.001 second and run the simulation upto $10^4$ seconds. For each realization the results are taken after ignoring the initial $10^3$  transients in order for the system to approach steady state.  The averages are taken over $10^4$ realizations.\\

\begin{figure*}[ht]
\includegraphics[scale=0.5]{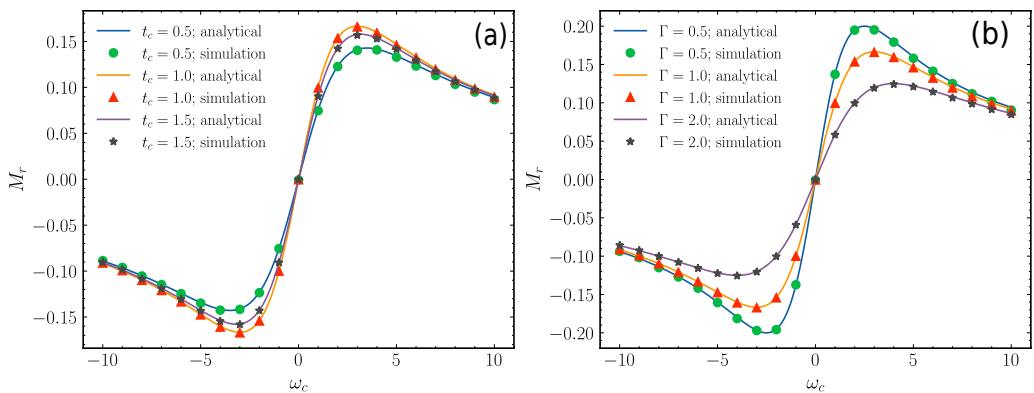}
\caption{$M_r$ [Eq.~\eqref{MagMomR}] is plotted with $\omega_c$ for different (a) $\Gamma$ and (b) for different $t_c$, keeping other parameters fixed as unity.}
\label{MTau}
\end{figure*}

From Eq.{\ref{MagMomR}}, one can also analyse the behaviour of $M_r$ with respect to $\Gamma$, $\omega_0$ and $\omega_c$. As $\Gamma \rightarrow \infty$, it is evident from the expression that $M_r=0$. It implies that in high viscous limit where inertia is negligibly small, the magnetic moment reduces to zero and it vanishes as $M_r\sim \Gamma^{-2}$.  Therefore the active particles can exhibit non-zero magnetic moment only when the dynamics of the system contains considerable inertia. Similarly from Eq.{\ref{MagMomR}}, it is evident that for large $\omega_0$, $M_r$ approaches towards zero as $M_r\sim \omega_0^{-4}$. Physically this occurs because, as $\omega_0$ increases the particles are constrained to move in smaller area and therefore the area of the current loop formed by the particles decreases. This eventually leads to zero magnetic moment. The dependence of $M_r$ on $\omega_c$ is non-monotonic. When $\omega_c=0$, $M_r=0$ and for small but finite $\omega_c$, $M_r$ increases with $\omega_c$ (for small $\omega_c$, $M_r\simeq \frac{|q|K_BT}{mc}t_c^2\omega_c$).  On the other hand, for large $\omega_c$, $M_r\simeq\left(\frac{|q|K_BT}{mc}\right)\frac{1}{\omega_c}$. Therefore, $M_r$ decreases with large $\omega_c$ and eventually it vanishes as $\omega_c\rightarrow \infty$. This non-monotonic dependence of $M_r$ on $\omega_c$ for different $\Gamma$ and $t_c$ is depicted in Fig.[\ref{MTau}], both from analytics and simulation.   

\section{CONCLUSION}
In this work, we consider a system of non-interacting charged particles, self-propelling in 2D, being suspended in a Maxwellian viscoelastic medium, with considerable inertia. They are confined in a harmonic trap and subject to Lorentz force due to externally applied constant magnetic field, perpendicular to the plane of their motion. Due to the imbalance between elastic as well as viscous dissipation (thanks to the medium) and active fluctuations (thanks to self-propulsion), the system goes out of equilibrium. The magnetic moment of the system can become non-zero and the system undergoes an interesting transition from diamagnetic phase to paramagnetic phase and vice-versa.  The transition depends on the interplay between the time scales involved in the dissipative processes and active fluctuations.

We have also determined how the magnetic moment of the system depends on the parameters like cyclotron frequency $\omega_c$ related to the strength of the magnetic field, natural frequency $\omega_0$ related to the strength of the harmonic trap and friction coefficient $\Gamma$. As $\Gamma\rightarrow\infty$, the magnetic moment vanishes, suggesting that the orbital magnetism of active viscoelastic suspension is exclusive for the active system with significant inertia.  

Interestingly, we find that even if the system remains under deep non-equilibrium conditions (in particular, when the particles are self-propelling with large persistence time scale), still it can have zero magnetic moment, as in case of equilibrium. 

It is important to explore how the results qualifies for relatively denser system with significant inter-particle interactions. Work in this direction is in progress. We believe all the aforementioned theoretical results are amenable to suitable experiments and they are important to  implement magnetic  control on the the dynamics of an active suspension by fine tuning external magnetic field.
 
\section{Acknowledgement}
M.M. and M.S. acknowledge the INSPIRE Faculty research grant (IFA 13 PH-66) by the Department of Science and Technology, Govt. of India. M.S. also acknowledge the UGC start-up grant and UGC Faculty recharge program (FRP-56055), UGC. A.S. thanks the start-up grant from UGC, UGCFRP and the Core Research Grant ( CRG/2019/001492) from DST, Govt. of India. We thank Sourabh Lahiri for critical reading of the manuscript.

\section{Author Contribution}
Both MS and AS have contributed equally to this work.


\bibliographystyle{unsrtnat}

\end{document}